\documentstyle[draft,epsf]{mn}

\def\1{\'\i}
\def\2{\c c}

\def\0{{$\clubsuit$}}

\title[Optical Monitoring  of Quasars]
       {Optical Monitoring of Quasars: I. Variability}
\author[A. Garcia, L. Sodr\'e Jr., F.J. Jablonski \& R.J. Terlevich]
       {Aurea Garcia$^{1}$\thanks{e-mail: aurea@iagusp.usp.br},  
    Laerte Sodr\'e Jr.$^{1}$\thanks{e-mail: laerte@iagusp.usp.br}, 
    Francisco J. Jablonski$^2$\thanks{e-mail: chico@das.inpe.br}
    \&  Roberto J. Terlevich$^{3}$\thanks{e-mail: rjt@ast.cam.ac.uk; 
visiting professor at Instituto Nacional de Astrof\1sica, Optica y 
Electr\'onica, Tonantzintla, Puebla, Mexico} \\ 
$^1$Departamento de Astronomia, Instituto Astron\^omico e Geof\'\i sico da 
USP, Av. Miguel Stefano 4200, 04301-904 S\~ao Paulo, Brazil \\
$^2$Departamento de Astrof\1sica INPE/MCT, CP151, 12201 S\~ao Jos\'e dos Campos, Brazil \\
$^3$Institute of Astronomy, Madingley Road, Cambridge CB3 OHA \\
}
\pagerange{\pageref{firstpage}--\pageref{lastpage}}

\newbox\grsign \setbox\grsign=\hbox{$>$} \newdimen\grdimen \grdimen=\ht\grsign
\newbox\simlessbox \newbox\simgreatbox
\setbox\simgreatbox=\hbox{\raise.5ex\hbox{$>$}\llap
     {\lower.5ex\hbox{$\sim$}}}\ht1=\grdimen\dp1=0pt
\setbox\simlessbox=\hbox{\raise.5ex\hbox{$<$}\llap
     {\lower.5ex\hbox{$\sim$}}}\ht2=\grdimen\dp2=0pt

\def\simless{\mathrel{\copy\simlessbox}}
\newbox\simppropto
\setbox\simppropto=\hbox{\raise.5ex\hbox{$\sim$}\llap
     {\lower.5ex\hbox{$\propto$}}}\ht2=\grdimen\dp2=0pt

\begin{document}

\label{firstpage}

\maketitle

\begin{abstract}
We present an analysis of quasar variability from data collected 
during a photometric monitoring of 50 objects carried out at 
CNPq/Laborat\'orio Nacional de Astrof\1sica, Brazil, between March 1993
and July 1996. A distinctive feature of this survey is its 
photometric accuracy, $\sim 0.02~ V$ mag, achieved through 
differential photometry with CCD detectors, what allows the
detection of faint levels of variability. We find that 
the relative variability, 
$\delta = \sigma / L$, observed in the $V$ band is  anti-correlated
with both luminosity and redshift, although we have no means of discovering
the dominant relation, given the strong coupling between luminosity and redshift for the objects in our sample.
We introduce a model for the dependence of quasar variability on frequency
that is consistent with multi-wavelength observations of the nuclear
variability of the Seyfert galaxy NGC 4151. We show that correcting the
observed variability for this effect slightly increases the significance of
the trends of variability with luminosity and redshift. Assuming that
variability depends only on the luminosity, we show that the
corrected variability is anti-correlated with luminosity and is in
good agreement with predictions of a simple Poissonian model.  The
energy derived for the hypothetical pulses, $\sim 10^{50}~{\rm erg}$, 
agrees well with those obtained in other studies. We also find
that the radio-loud objects in our sample tend to be more variable than the 
radio-quiet ones, for all luminosities and redshifts.

\end{abstract}


\section{Introduction}
Although variability is a well known property of quasars and Seyfert 1
galaxies, many of its properties are still subject of debate. 
As noticed by Pica \& Smith (1983), studying the dependence of
the variability on the luminosity may allow to verify
whether the variability is Poissonian or not, that is, whether the source is
intrinsically multiple (like compact supernova remnants evolving in a 
nuclear starburst; e.g., Terlevich et al. 1992) or may be considered 
as a single coherent source (like an accretion disk around a massive black 
hole; e.g., Rees 1984). 

A main prediction of the sub-units model, as sometimes the Poissonian model is 
called, is that the variability (measured in magnitudes) should be anti-correlated 
with the source luminosity. Such a behaviour was observed in some early studies
(e.g., Uomoto, Wills \& Wills 1976; Pica \& Smith 1983) and, more recently, in the analysis 
of the samples SA 94 (Cristiani, Vio \& Andreani 1990) and SA 57 (Trevese et al. 1994), 
as well as in the study of the SGP sample (Hook et al. 1994). These three samples 
were jointly revisited by Cristiani et al. (1996), that confirmed the anti-correlation 
obtained previously. However, most of these studies obtained a logarithmic slope for 
the relation between variability and luminosity shallower than the value -0.5
that is expected in the simple Poissonian model (see section 4.1). 
The same result was achieved by Paltani \& Courvoisier (1997), 
who analyzed the ultraviolet continuum of all Seyfert 1 galaxies,
radio-quiet quasars, and low polarization radio-loud quasars 
contained in the IUE database. However, other studies have indeed shown evidence 
for a Poissonian behaviour. For instance, Cid Fernandes, Aretxaga \& Terlevich 
(1996) analyzed the variability of the SGP sample with respect to
the luminosity and redshift, obtaining a logarithmic slope consistent with the 
expected value of -0.5. Also, Aretxaga, Cid Fernandes \& Terlevich (1997) 
have shown that the variability of the SGP sample can be well-reproduced 
by a random superposition of discrete events. 
 
However, even the very existence of an anti-correlation between variability 
and luminosity has been disputed. For instance, Lloyd (1984), 
Cimatti, Zamorani \& Marano (1993), and Netzer et al. (1996) have found no dependence of 
variability on absolute magnitude. Such a controversy probably arises
from the different characteristics of the samples studied and
procedures used to analyze them. Only recently CCDs have been introduced
in quasar monitoring campaigns, improving the accuracy of data and allowing
the detection of low levels of variability (see, for instance,
Borgeest \& Schramm 1994, Netzer et al. 1996 and Giveon et al. 1999). Besides, most of the monitored samples
present a strong coupling between luminosity and redshift, typical of flux 
limited samples. Undoubtedly, this is a major problem of the available data sets,
because it prevents one of disentangling the real dependence of the variability
on these two parameters. In particular, Cristiani et al. (1996) divided their
data in slices in luminosity and redshift, and concluded that there was a 
significant positive correlation of the variability with redshift. 
This trend had already been detected by Giallongo, Trevese \& Vagnetti
(1991), and was interpreted as resultant of a variability vs.
wavelength dependence (see below). Cid Fernandes, Aretxaga \& Terlevich (1996)
also supported this hypothesis to explain the positive correlation 
between variability and redshift found in their fits.

The increase of the variability with increasing frequency has been observed 
by many authors (e.g.,  Cutri et al. 1985; Edelson, Krolik \& Pike 1990; 
Kinney et al. 1991; Paltani \& Courvoisier 1994; Cristiani et al. 1997;
Paltani, Courvoisier \& Walter 1998). A good example of such behaviour 
in a low luminosity AGN can be seen in Edelson et al. (1996), that contains 
multi-wavelength variability data of NGC 4151, monitored through the AGN Watch 
consortium. The parameterization of this trend is 
of serious concern, because observations in a given photometric
band sample different parts of the rest-frame spectrum of quasars at
different redshifts. Of course, such effect should be taken in to account in 
the analysis of the variability of large samples of quasars.
If well understood, this effect can be used to constrain models for
the nature of quasars and other forms of AGN.  

In this paper we present the first results of a monitoring of 50
quasars that has been conducted at the CNPq/Laborat\'orio Nacional 
de Astrof\1sica, in Brazil, since 1993. The observations discussed 
here were done with a 0.6m telescope in the $V$ 
band during the period between March 1993 and July 1996. 
The observational strategy is based on differential photometry with CCD 
detectors. This provides good photometric accuracy, necessary to detect 
very low levels of variability.
These observations will be used to address several issues regarding 
the optical variability of quasars. In particular, we discuss here the
relation of variability with luminosity, including a model
for the dependence of variability on frequency. We will also re-examine 
whether a simple Poissonian model is consistent with the observations,
using a robust fitting technique. We will also compare the variability 
properties of the radio-loud and radio-quiet objects present in our sample.
In another paper, we will present an analysis of the structure function
of the light-curves, useful to estimate the variability time-scales of quasars
(Garcia et al., in preparation).

This paper is organized as follows. Section 2 presents our sample and a
description of the observations and data reduction, as well as the
light curves of the quasars. In section 3 we present our estimator of
quasar variability and a simple model for correcting the observations
of the dependence of variability with frequency. In section 4 we present 
the simple Poissonian model and
examine whether it is able to explain our observations, taking into account
the correction for the variability vs. frequency effect. 
In section 5 we compare
the variability observed in the radio-loud and radio-quiet subsamples and,
finally, we summarize our conclusions in section 6.

\section{\bf The Data}

\subsection{\bf The Sample}

The sample of 50 quasars discussed here was drawn from table 1 of  
V\'eron-Cetty \& V\'eron (1987) catalog. All quasars have
redshifts larger than 0.15 and V-band apparent magnitudes brighter than 
16.5 mag 
in the catalog. The former criterion aimed to minimize the light contribution 
from the host galaxy, whereas the latter ensured that the sample was bright 
enough to make the measurement errors small, less than 0.05 mag. 
Classifying objects with radio emission at 5 GHz larger than 
10$^{25}$ W Hz$^{-1}$ as radio-loud (Kellermann et al. 1994), there are 35 
radio-loud and 15 radio-quiet quasars in the sample. Among the
radio-loud objects, one BL Lac (PKS 0537-441) was accidentally selected,
because it appeared in the 1987 version of the V\'eron-Cetty \& V\'eron 
catalog as a ``normal" quasar, an error corrected in more recent versions
of that catalog.

\begin{figure*}
\centerline{\epsfxsize= 12cm \epsfbox{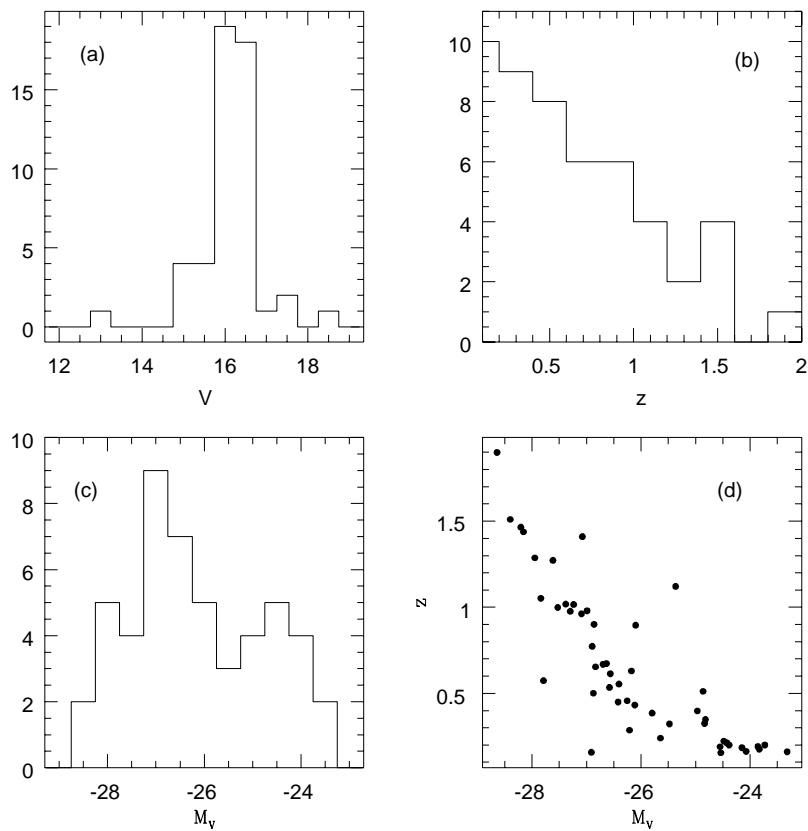}}
\caption{Distribution of the sample in (a) apparent $V$ magnitudes,
(b) redshift, and (c) absolute $V$ magnitudes. 
It is also shown the Hubble diagram of the sample (panel 1-d). }
\label{fig4}
\end{figure*}

The sample is presented in Table 1. The columns of the table give, for each 
object, its name, the equatorial coordinates $\alpha$ and $\delta$ (J2000), 
the radio class R (L: radio-loud, Q: radio-quiet), the redshift $z$, the mean
apparent and absolute $V$ magnitudes ($V$ and $M_V$), the number of observed 
epochs ($n_e$), the observer and rest-frame time coverage 
($\Delta t_{obs}, \Delta t_{rest}$, in years),  and the observed relative 
variability ($\delta_{obs}=(\sigma/L)_{obs}$) and the mean relative 
measurement error  
($\delta_\epsilon =\epsilon/L $) in the $V$ band. Three objects in the sample 
are high polarization quasars (see table 1) according to V\'eron-Cetty 
\& V\'eron (1993) or NED 
{\footnote{The NASA/IPAC Extragalactic Database is operated by the Jet Propulsion
Laboratory, California Institute of Technology, under contract with the
National Aeronautics and Space Administration.}}; 
one of them is the
blazar PKS 0537-441. The apparent $V$ magnitudes in table
1 were computed from our photometric calibration for all but 5 quasars (see
table 1) for which we adopted magnitudes
from V\'eron-Cetty \& V\'eron (1993; see below).

The absolute magnitudes in table 1 were computed assuming
that the optical-UV continuum of quasars may be described by a power-law 
$f_\nu \propto \nu^{\alpha}$ with spectral index $\alpha$=$-$0.3
(Francis 1996, Peterson 1997). Besides the $k$-correction, we
have also included an empirical correction 
that takes into account the presence of emission lines and the Lyman forest 
on the measured continuum (V\'eron-Cetty \& V\'eron 1993). 
The reddening was corrected from the $A_B$ values given by NED,
assuming that $A_V=0.77~A_B$ (Allen 1976).
Luminosity distances were computed assuming H$_0$=50 km s$^{-1}$
Mpc$^{-1}$, $\Omega_0$=1, and $\Lambda=0$.

Figure 1 shows same features of the sample discussed here. Figure 1-a indicates
that most of the objects in the sample have indeed mean apparent magnitudes
around $V$=16 (median $V=16.17$). The redshift distribution is 
presented in figure 1-b (median $z=0.54$). Figure 1-c presents
the $M_V$ distribution (median $M_V=-26.4$).
The peak around $M_V \simeq -24.5$ is due to 
radio-quiet, low redshift objects. The Hubble diagram, i.e., $M_V \times z$,
is shown in figure 1-d. Note how strongly luminosity and redshift are
correlated in this sample. This happens because, although the quasars in the 
sample span a large range of redshifts, most of them have mean apparent $V$
magnitudes around 16. 

\begin{table*}
\begin{minipage}{17.5cm}
\caption{Description of the sample and observational information.}
\footnotesize{
\begin{center}
\begin{tabular}{lllcccccccccc}

Quasar & $\alpha$ (2000) & $\delta$ (2000) & R$^{(2)}$ & z & $V$ & $M_V$ & $n_e$ & $\Delta t_{obs}$ &
$\Delta t_{rest}$ & $\delta_{obs}$ & $\delta_\epsilon$ \\
& & & & & & & & (yr) & (yr) & ($V$ band) & ($V$ band) \\
UM 18           &  00 05 20.2 & +05 24 12 & L & 1.899  &   16.29  &  -28.64  &    7  &  2.74  &  0.95  &  0.038  &  0.022   \\
PKS 0003+15     &  00 05 59.2 & +16 09 50 & L & 0.450  &   15.62  &  -26.42  &    6  &  2.74  &  1.89  &  0.103  &  0.016   \\
PKS 0005-239    &  00 07 56.1 & --23 41 17 & L & 1.410  &   17.32  &  -27.08  &    5  &  2.74  &  1.138 &  0.017  &  0.018   \\
PG 0043+039     &  00 45 47.2 & +04 10 24 & Q & 0.385  &   15.89$^{(3)}$  &  -25.80  &    9  &  2.82  &  2.03  &  0.056  &  0.047   \\
Mark 1014       &  01 59 50.1 & +00 23 43 & Q & 0.163  &   15.69$^{(3)}$  &  -24.07  &    6  &  2.02  &  1.74  &  0.084  &  0.048   \\
3C 57           &  02 01 57.1 & --11 32 32 & L & 0.669  &   16.14  &  -26.70  &    6  &  1.95  &  1.17  &  0.060  &  0.017   \\
PKS 0232-04     &  02 35 07.2 & --04 02 05 & L & 1.438  &   16.26  &  -28.16  &    7  &  1.94  &  0.80  &  0.060  &  0.019   \\
PKS 0312-77     &  03 11 55.3 & --76 51 51 & L & 0.223  &   16.10$^{(3)}$  &  -24.48  &    8  &  3.40  &  2.78  &  0.122  &  0.049   \\
3C 95           &  03 51 28.5 & --14 29 08 & L & 0.614  &   16.11  &  -26.56  &    7  &  1.94  &  1.20  &  0.258  &  0.016   \\
3C 94           &  03 52 30.5 & --07 11 01 & L & 0.962  &   16.45  &  -27.09  &    7  &  1.94  &  0.99  &  0.023  &  0.015   \\
PKS 0355-483    &  03 57 22.0 & --48 12 16 & L & 1.016  &   16.38$^{(3)}$  &  -27.24  &    8  &  3.25  &  1.61  &  0.027  &  0.047   \\
PKS 0405-12     &  04 07 48.3 & --12 11 35 & L & 0.574  &   14.78  &  -27.79  &    7  &  1.94  &  1.23  &  0.074  &  0.015   \\
3C 110          &  04 17 16.7 & --05 53 45 & L & 0.773  &   16.20  &  -26.90  &    5  &  1.94  &  1.10  &  0.028  &  0.016   \\
PKS 0454-22     &  04 56 08.9 & --21 59 09 & L & 0.534  &   15.80  &  -26.58  &    5  &  2.18  &  1.42  &  0.106  &  0.017   \\
PKS 0537-441$^{(1)}$   &  05 38 50.3 & --44 05 09 & L & 0.896  &   16.48$^{(3)}$  &  -26.10  &    8  &  3.32  &  1.75  &  1.483  &  0.124   \\
PKS 0637-75     &  06 35 46.5 & --75 16 17 & L & 0.654  &   16.09  &  -26.84  &    9  &  3.47  &  2.10  &  0.188  &  0.014   \\
PKS 0736+01$^{(1)}$    &  07 39 17.9 & +01 37 05 & L & 0.191  &   16.31  &  -23.86  &    8  &  2.41  &  2.02  &  0.547  &  0.033   \\
PKS 0743-67     &  07 43 31.6 & --67 26 26 & L & 1.510  &   16.34  &  -28.40  &    8  &  3.47  &  1.38  &  0.124  &  0.030   \\
PKS 0837-12     &  08 39 50.5 & --12 14 34 & L & 0.200  &   16.52  &  -23.73  &    9  &  2.60  &  2.16  &  0.212  &  0.012   \\
DW 0839+18      &  08 42 05.0 & +18 35 41 & L & 1.272  &   16.54  &  -27.62  &    9  &  2.41  &  1.06  &  0.153  &  0.014   \\
PG 0923+201     &  09 25 54.6 & +19 54 05 & Q & 0.190  &   15.58  &  -24.55  &   10  &  2.60  &  2.18  &  0.093  &  0.012   \\
PKS 0925-203    &  09 27 51.7 & --20 34 51 & L & 0.348  &   16.65  &  -24.82  &   10  &  2.60  &  1.92  &  0.100  &  0.018   \\
PG 1001+05      &  10 04 20.1 & +05 13 01 & Q & 0.161  &   16.41  &  -23.32  &   10  &  2.59  &  2.23  &  0.066  &  0.015   \\
PKS 1004+13     &  10 07 26.1 & +12 48 57 & L & 0.240  &   14.98  &  -25.64  &   10  &  3.14  &  2.53  &  0.144  &  0.012   \\
PG 1008+133     &  10 11 10.8 & +13 04 13 & Q & 1.287  &   16.27  &  -27.95  &    7  &  3.14  &  1.37  &  0.027  &  0.014   \\
PG 1012+00      &  10 14 54.8 & +00 33 37 & Q & 0.185  &   15.90  &  -24.15  &    8  &  3.14  &  2.65  &  0.074  &  0.015   \\
PG 1151+117     &  11 53 49.3 & +11 28 31 & Q & 0.176  &   16.10  &  -23.83  &    9  &  3.14  &  2.67  &  0.084  &  0.012   \\
3C 273          &  12 29 06.7 & +02 03 09 & L & 0.158  &   12.76  &  -26.91  &   10  &  3.14  &  2.71  &  0.121  &  0.016   \\
PG 1254+047     &  12 56 59.9 & +04 27 35 & Q & 1.018  &   16.24  &  -27.38  &   10  &  3.14  &  1.56  &  0.057  &  0.010   \\
PKS 1302-102    &  13 05 32.9 & --10 33 20 & L & 0.286  &   14.82  &  -26.21  &   11  &  3.22  &  2.50  &  0.089  &  0.217   \\
PG 1307+085     &  13 09 47.0 & +08 19 51 & Q & 0.155  &   15.12  &  -24.54  &    6  &  0.41  &  0.35  &  0.063  &  0.173   \\
PG 1333+176     &  13 36 02.0 & +17 25 14 & Q & 0.554  &   16.06  &  -26.41  &   10  &  2.09  &  1.34  &  0.029  &  0.117   \\
1E 1352+1820    &  13 54 35.6 & +18 05 19 & Q & 0.977  &   16.22  &  -27.30  &    5  &  0.33  &  0.17  &  0.011  &  0.043   \\
PG 1352+011     &  13 54 59.8 & +00 52 50 & Q & 1.121  &   18.51  &  -25.36  &    8  &  3.22  &  1.52  &  0.066  &  0.059   \\
PKS 1912-549    &  19 16 39.2 & --54 54 47 & L & 0.398  &   16.76  &  -24.97  &   17  &  3.16  &  2.26  &  0.187  &  0.022   \\
PKS 2021-330    &  20 24 35.5 & --32 53 36 & L & 1.465  &   16.30  &  -28.20  &   16  &  3.82  &  1.55  &  0.066  &  0.013   \\
PG 2112+059     &  21 14 52.6 & +06 07 44 & Q & 0.457  &   15.84  &  -26.26  &   14  &  3.22  &  2.21  &  0.035  &  0.016   \\
PKS 2115-30     &  21 18 10.5 & --30 19 10 & L & 0.980  &   16.61  &  -26.99  &   10  &  2.82  &  1.42  &  0.089  &  0.017   \\
PKS 2128-12     &  21 31 35.2 & --12 07 04 & L & 0.501  &   15.40  &  -26.87  &   10  &  2.82  &  1.88  &  0.068  &  0.031   \\
PKS 2135-14     &  21 37 45.1 & --14 32 55 & L & 0.200  &   15.83  &  -24.39  &    9  &  2.09  &  1.74  &  0.237  &  0.040   \\
OX 169          &  21 43 35.5 & +17 43 49 & L & 0.213  &   16.04  &  -24.43  &   10  &  2.67  &  2.20  &  0.140  &  0.032   \\
PKS 2145+06     &  21 48 05.4 & +06 57 40 & L & 0.999  &   16.07  &  -27.53  &    9  &  2.67  &  1.34  &  0.144  &  0.023   \\
PKS 2216-03     &  22 18 52.0 & --03 35 36 & L & 0.901  &   16.52  &  -26.86  &    7  &  2.09  &  1.10  &  0.049  &  0.068   \\
PG 2233+134     &  22 36 07.7 & +13 43 56 & Q & 0.325  &   16.50  &  -24.84  &   11  &  3.22  &  2.43  &  0.041  &  0.056   \\
PKS 2243-123$^{(1)}$   &  22 46 18.2 & --12 06 51 & L & 0.630  &   16.61  &  -26.18  &    9  &  2.82  &  1.73  &  0.106  &  0.084   \\
PKS 2251+11     &  22 54 10.4 & +11 36 40 & L & 0.323  &   15.85  &  -25.48  &    8  &  2.67  &  2.02  &  0.050  &  0.090   \\
PKS 2300-683    &  23 03 43.6 & --68 07 37 & L & 0.512  &   17.40  &  -24.86  &   10  &  3.82  &  2.53  &  0.120  &  0.046   \\
PG 2302+029     &  23 04 44.9 & +03 11 47 & Q & 1.052  &   15.95  &  -27.84  &    8  &  2.75  &  1.34  &  0.053  &  0.011   \\
4C 09.72        &  23 11 17.7 & +10 08 18 & L & 0.432  &   15.82  &  -26.12  &    6  &  2.75  &  1.92  &  0.133  &  0.079   \\
PKS 2344+09     &  23 46 36.7 & +09 30 46 & L & 0.673  &   16.28  &  -26.64  &    7  &  2.75  &  1.64  &  0.056  &  0.014   \\
\end{tabular}
\end{center}}
Notes: (1) high polarization quasars; (2) radio class: radio-loud (L) or radio-quiet (Q);
(3) objects for which the $V$ magnitudes were taken from V\'eron-Cetty \& 
V\'eron (1993).
\end{minipage}
\end{table*}

\subsection{\bf Observations and Data Reduction}

The observations were carried out at the 0.6m Boller \& Chivens telescope 
located at the Observat\'orio do Pico dos Dias, operated by the
CNPq/Laborat\'orio Nacional de Astrof\1sica, in Brazil, during the period 
from March 1993 to July 1996. We used basically two observational setups: one 
with the imaging camera coupled directly to a 770 $\times$ 1152 pixels EEV CCD,
and the other with a focal reducer plus a 385 $\times$ 578 pixels EEV 
CCD, in order to have a field of view of about $7^\prime \times 11^\prime$ 
in both cases. This field is large enough to allow the simultaneous
observation of the quasar and some reference stars for the differential 
photometry.
All observations were made in the $V$ band, with two combinations of filters 
(GG495/2+BG18/2 or GG495+BG39/3, Bessel 1990); we have verified that the
differential magnitudes obtained with these two filters are essentially the
same. The quasar fields were
observed with two successive exposures, each of them of 5 minutes,
to minimize cosmic ray contamination. This integration time allow us to 
achieve a photometric accuracy of a few hundredths of magnitude for most of
the objects in the sample.

The data were reduced using IRAF 
{\footnote {IRAF is distributed by National
Optical Astronomy Observatories, which is operated by the Association of
Universities for Research in Astronomy, Inc., under contract with the
National Science Foundation.}}
standard packages for 
CCD reduction and photometry. The images were first bias and flat-field 
corrected using the package {\it ccdred}.
The magnitudes were measured from aperture photometry through the 
{\it phot} task in the {\it daophot} package. 
In each epoch (corresponding to each observational run) we extracted 
magnitudes within several aperture radii, and then chose the radius 
which minimized the photometric errors.

Differential photometry is done with respect to reference stars observed
in the same field as the quasars. The procedure we devised to select these
stars for each field is as follows. First, we selected the brightest
isolated stellar sources as possible reference stars for that field.
After, we computed magnitude differences between pairs of stars,
selecting as reference the star that presented the lowest magnitude
dispersion relative to the others during all epochs.

By analyzing multiple observations of several fields, we detected a median
underestimation of the photometric errors given by {\it phot} by a factor 
of 1.73, similar to that obtained by Gopal-Krishna, Sagar \& Wiita (1995), of 1.75. 
After intensive study of the possible causes of this underestimation, we
concluded that it could be originated by random fluctuations of the PSF over
the image, probably induced by flat field errors. In fact, large scale 
fluctuations produced by the dome flat fields used in the reduction may 
well increase the errors to the observed values. 
To overcome this problem, we decided to adopt a minimum error of 0.01 mag 
for each individual magnitude extraction, since this is the expected
photometric error of a $V=$16 object with the observational setup used
in the observations.

We performed magnitude calibrations by observing standard stars. These
observations were done in 7 nights, during runs in March, May, July and 
August of 1996, and May 1997. The standard stars were selected from Graham 
(1982) list, and observed for different air masses every two hours during the 
calibration nights. As the observations were conducted at only one band ($V$), 
we have taken into account only the extinction coefficient of first order, 
associated with the airmass, to determine the apparent magnitudes of the 
reference star in each quasar field. These observations allowed us to 
calibrate the magnitudes of 45 quasars. The remaining 5 quasars not 
observed during the calibration nights (see Table 1)
were assumed to have average apparent 
magnitudes equal to those reported in the V\'eron-Cetty \& V\'eron (1993) 
catalog. 
The photometric errors achieved in our survey are relatively
low  (the median value of $\delta_\epsilon$ is 0.019),
what allows the detection of low levels of variability.

All the observations made within 5 days were 
grouped in a single observation- an epoch. The entry $n_e$ in Table 1 
refers to these grouped observations. The median number of epochs per quasar
is 8, and the median time interval covered by the observations is 2.7 and 1.7
years, in the observer and quasar rest-frame, respectively. Table 2 summarizes
some features of the sample, presenting the median values of most of the 
entries in Table 1, for all objects of the sample as well as for the
radio-loud and radio-quiet sub-samples.

\begin{table}
\caption{Median values of some quantities of the sample}
\begin{tabular}{lrrr}
 & all sample & radio-loud & radio-quiet \\
 number & 50 & 35 & 15 \\
 $z$ &   0.544 &   .614 &  .385 \\
 $V$ &  16.17 & 16.26 & 16.06 \\
 $M_V$ &  -26.42 & -26.64 & -25.36 \\
 $n_e$ &   8 &   8 &   9 \\
 $\Delta t_{obs}$ (yr) &   2.74 &  2.74 &  2.82 \\
 $\Delta t_{rest}$ (yr) &    1.74 &  1.73 &  1.74 \\
 $\delta_{obs}$ &    .079 &  .106 &  .057 \\
 $\delta_\epsilon$ &    .019 &  .019 &  .016 \\
\end{tabular}
\end{table}

\subsection{Light Curves}

The differential light curves of the quasars in our sample are shown in 
figures 2 to 8. Each data point ($\Delta V \equiv$ quasar magnitude minus 
reference star magnitude) corresponds to the average of all the observations
collected in each epoch, and the error bars are the corresponding photometric 
errors. 

The observed light curves present a rich variety of forms.
Some objects show a monotonical increase or decrease in luminosity 
(e.g. 3C 273, PKS 2128-12, 4C 09.72), while others seem to exhibit
a pulse-like pattern (PKS 1912-549). However, the time-scale covered by the
observations (and also the sampling rate) is not large enough to 
allow an appropriate discussion of the form of the light curves. Since we
continue monitoring this sample, we postpone this discussion for a future
paper, where we will discuss the data collected over a larger time interval. 

\section{Measurements of Variability}

\begin{figure*}
\centerline{\epsfxsize= 19.0cm \epsfbox{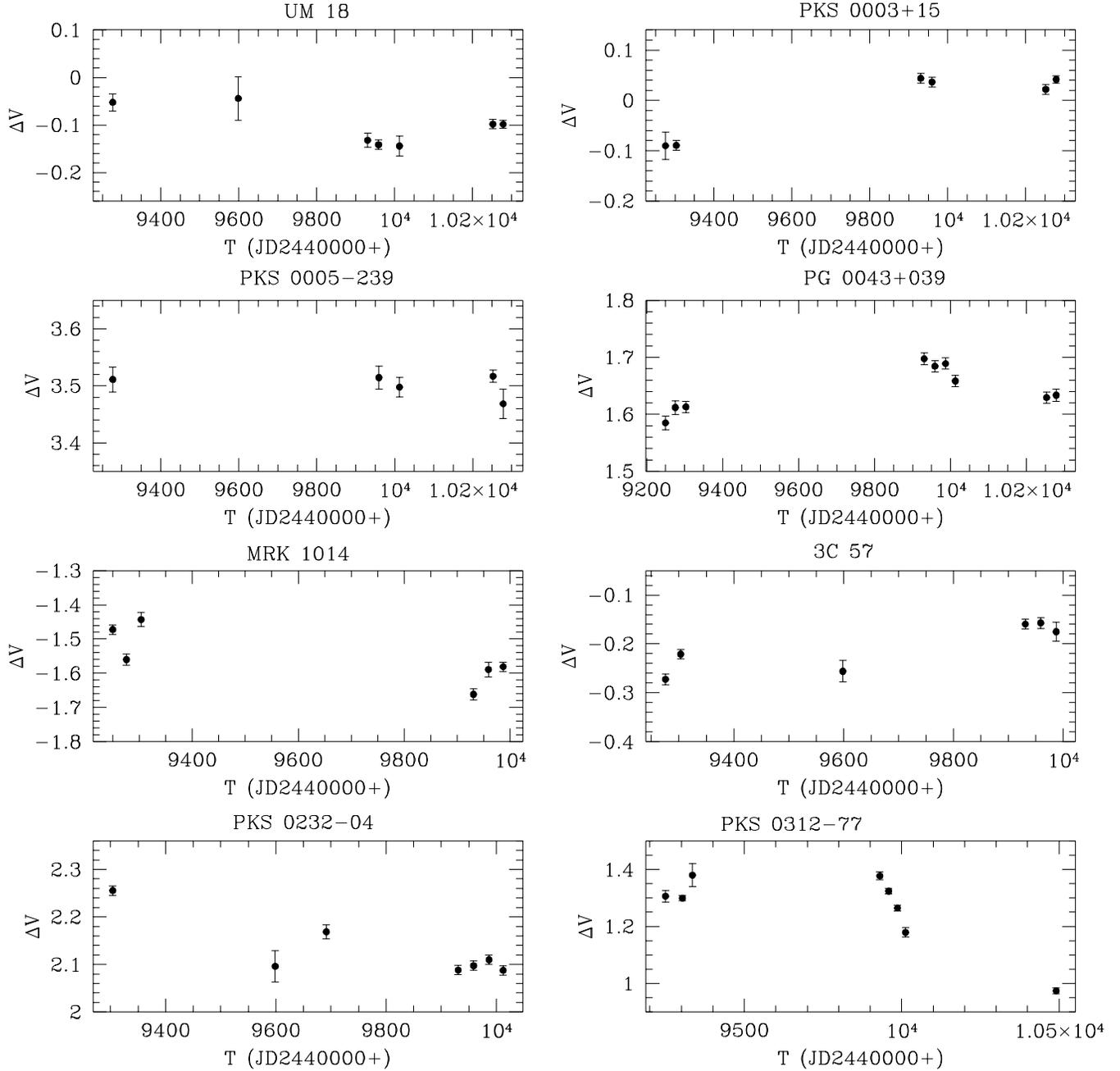}}
\caption{Differential light curves of the sample. $\Delta$V is the difference
between the magnitudes of the quasar and the corresponding reference star.}
\end{figure*}

\begin{figure*}
\centerline{\epsfxsize= 19.0cm \epsfbox{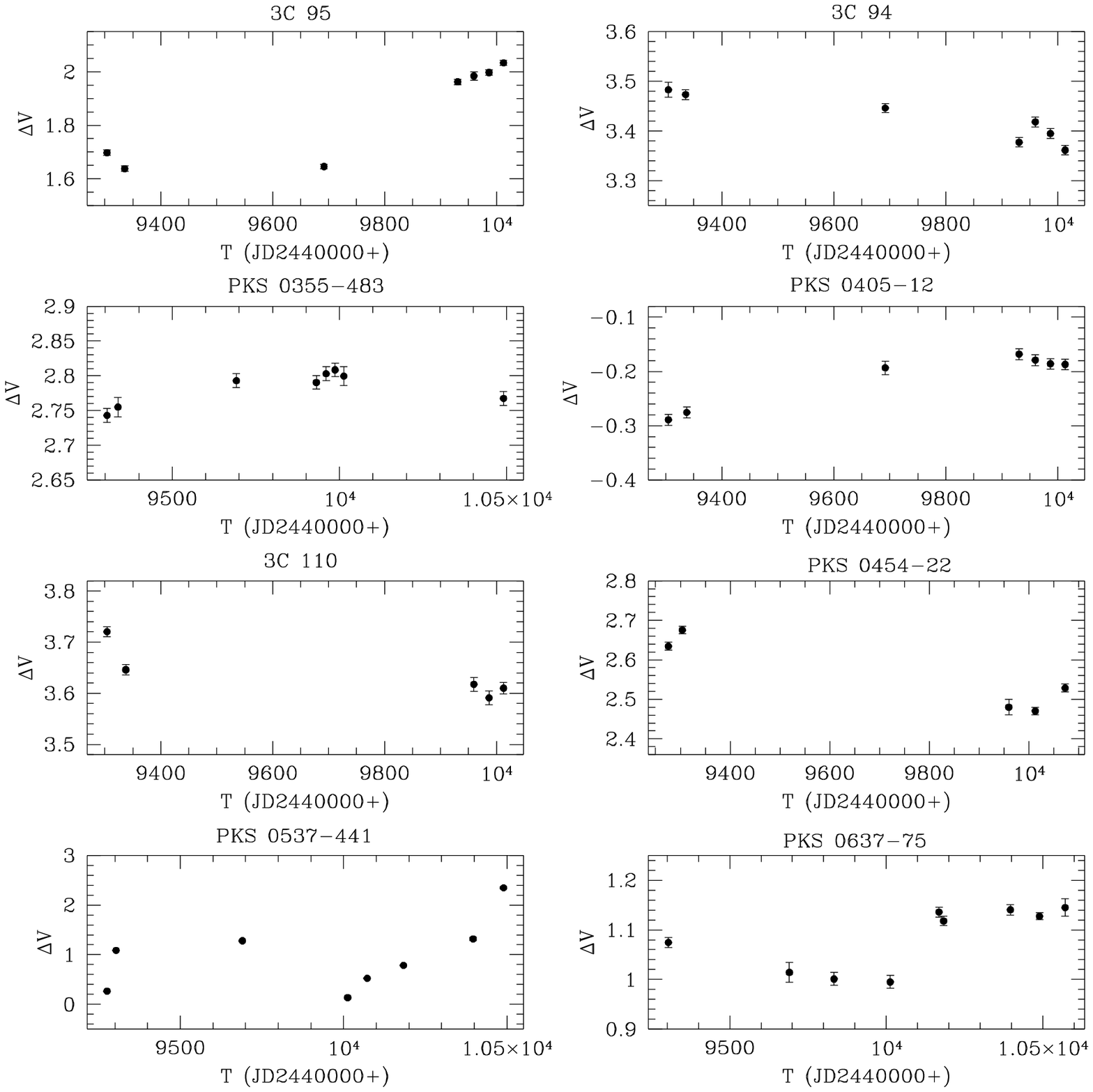}}
\caption{Differential light curves of the sample - cont.}
\end{figure*}

\begin{figure*}
\centerline{\epsfxsize= 19.0cm \epsfbox{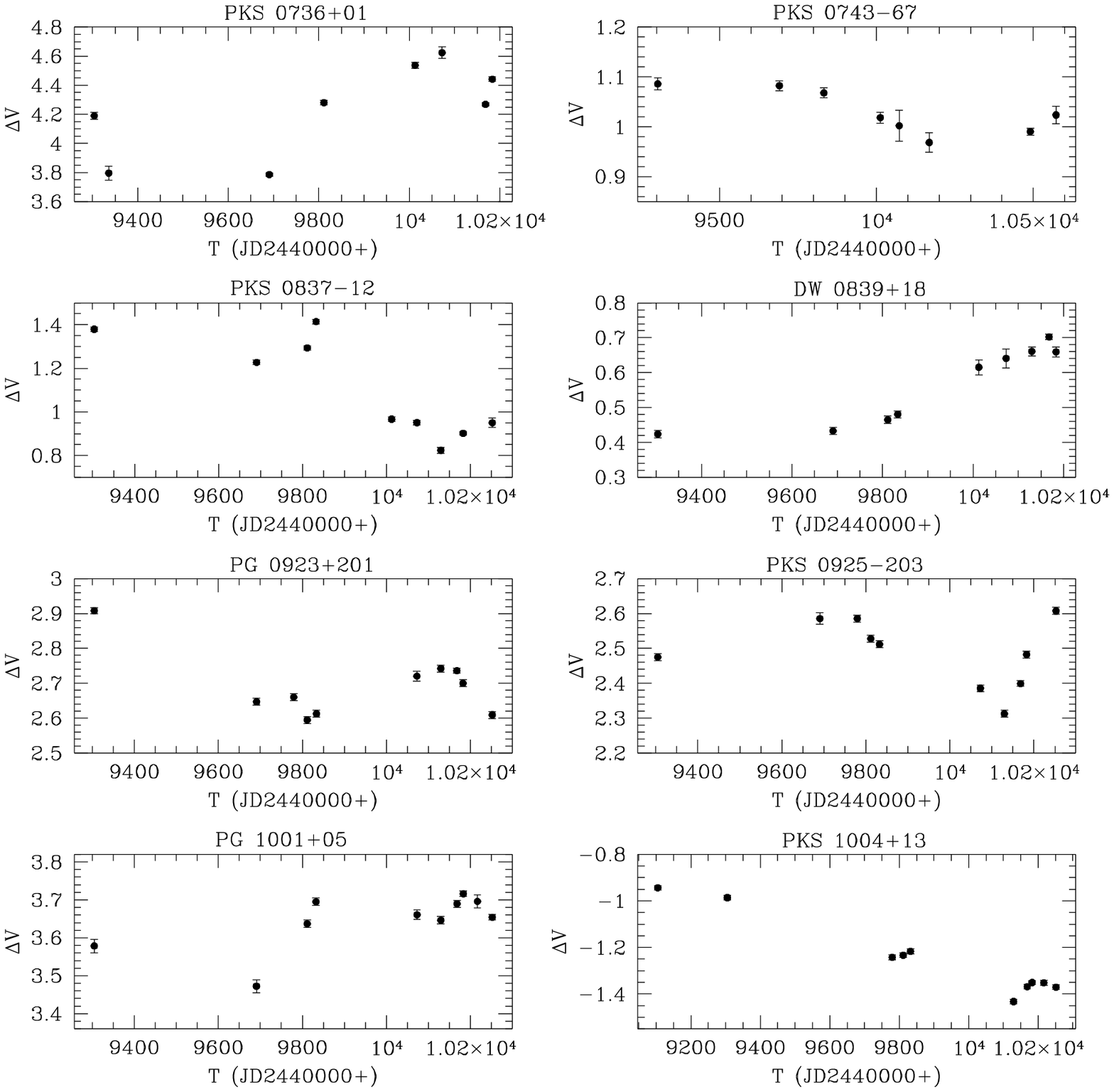}}
\caption{Differential light curves of the sample - cont.}
\end{figure*}

\begin{figure*}
\centerline{\epsfxsize= 19.0cm \epsfbox{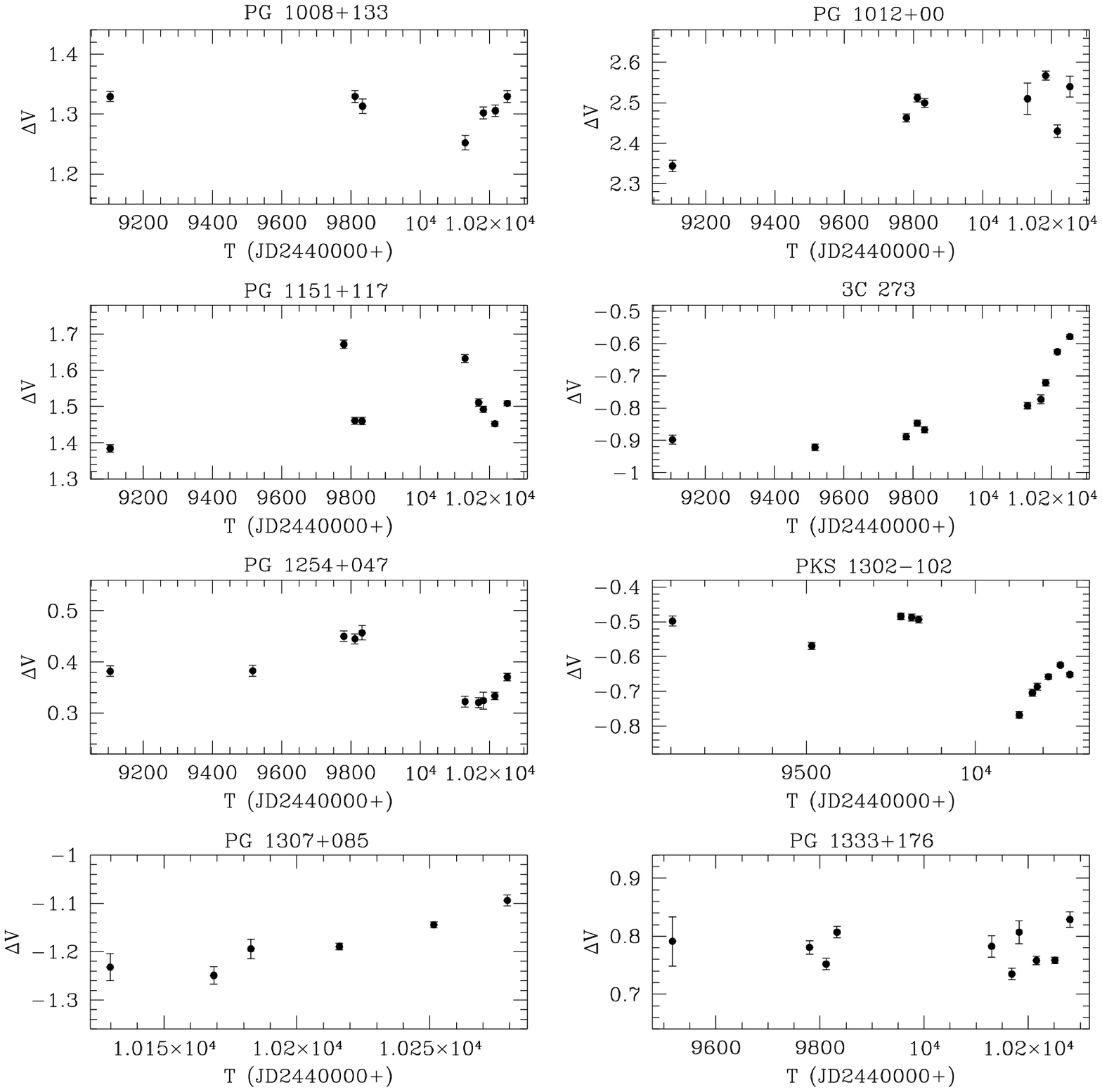}}
\caption{Differential light curves of the sample - cont.}
\end{figure*}

\begin{figure*}
\centerline{\epsfxsize= 19.0cm \epsfbox{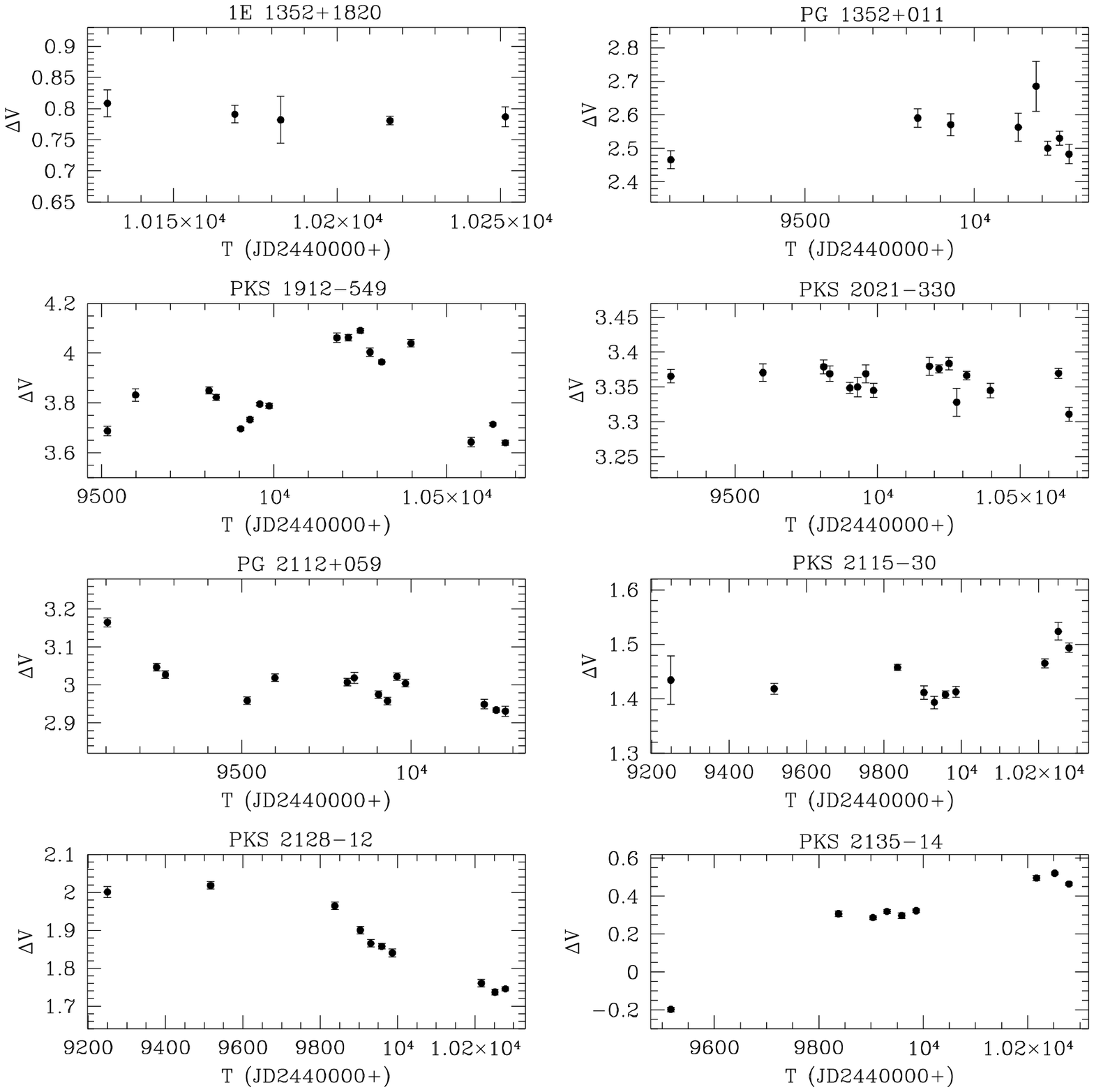}}
\caption{Differential light curves of the sample - cont.}
\end{figure*}

\begin{figure*}
\centerline{\epsfxsize= 19.0cm \epsfbox{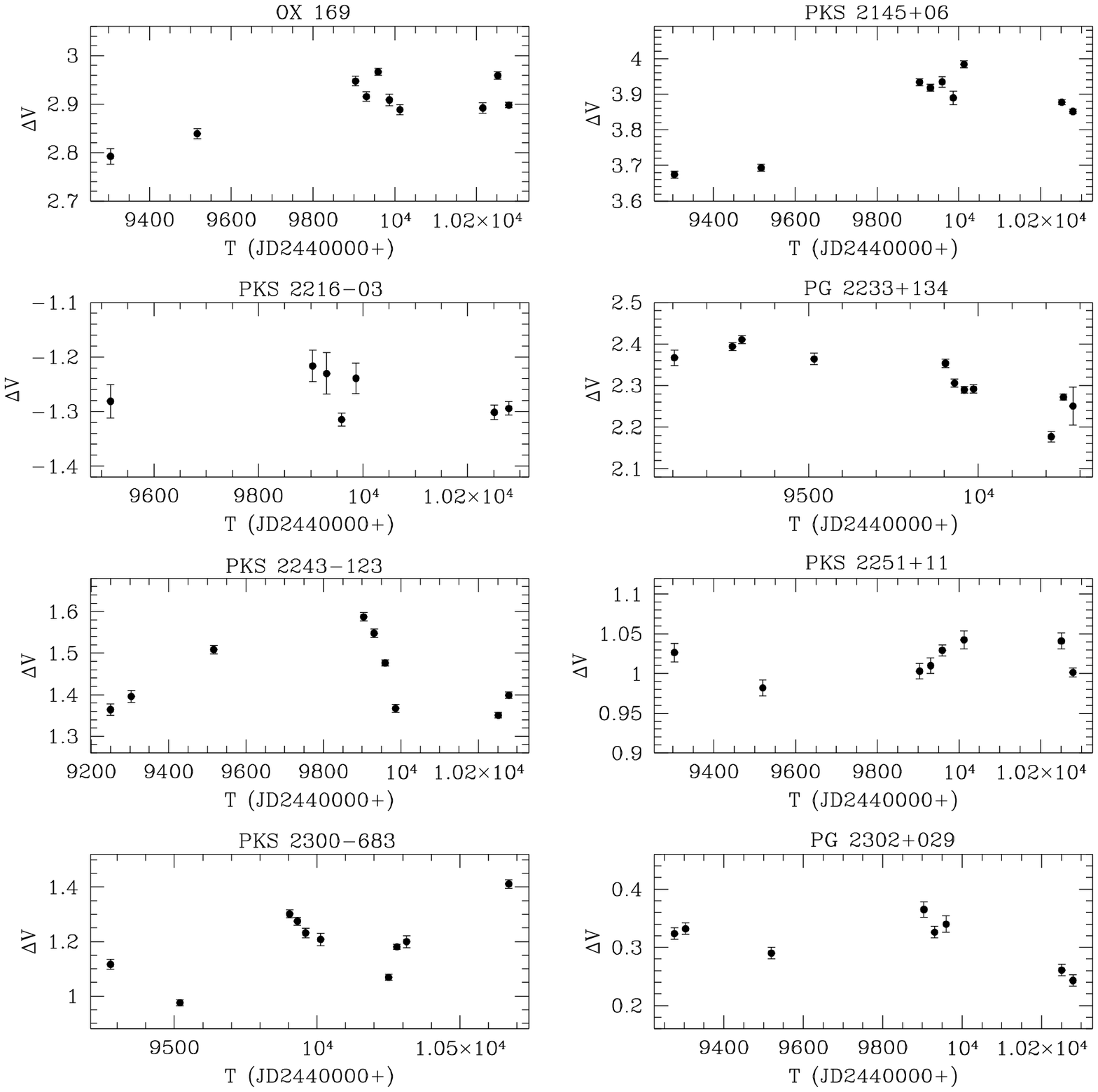}}
\caption{Differential light curves of the sample - cont.}
\end{figure*}

\begin{figure*}
\centerline{\epsfxsize= 19.0cm \epsfbox{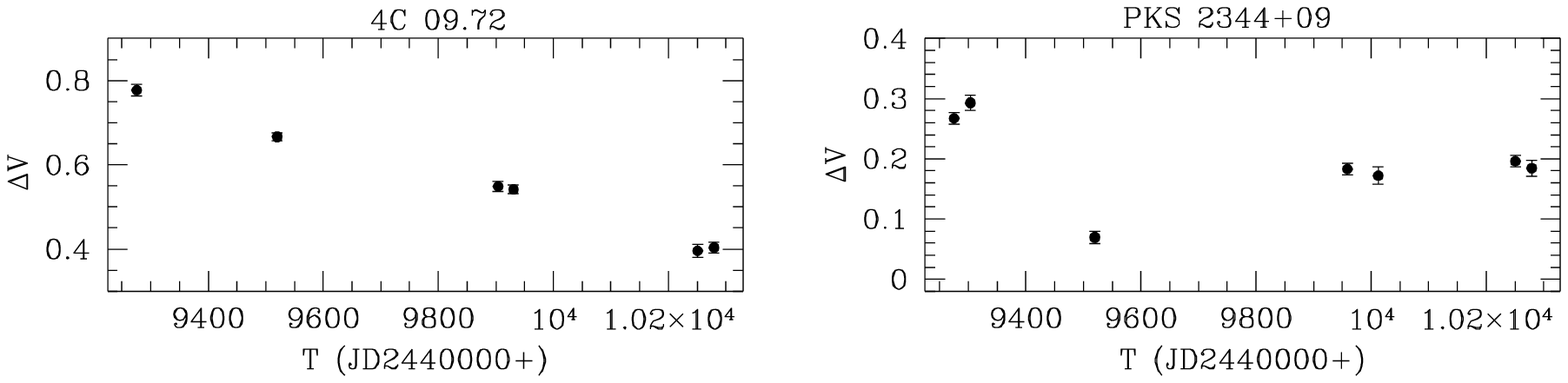}}
\vskip -14.cm
\caption{Differential light curves of the sample - cont.}
\end{figure*}

\subsection{Variability Index}

The variability analysis in this paper will be done in terms of
luminosity, instead of magnitudes. This procedure makes the comparison with
model predictions more direct.
We estimate the variability of a quasar with a ``robust" index 
similar to that used by Hook et al. (1994):
\begin{equation}
\sigma_{obs} = \frac{1}{n_e-1} \sqrt{\frac{\pi}{2}}\sum_{i=1}^{n_e} 
\left|L_i -L\right|
\end{equation}
where $n_e$ is the number of epochs, $L_i$ is the quasar luminosity
in the $V$ band  
at the $i-$th epoch, and $L$ is the mean quasar luminosity. 
The factor $\sqrt{\pi/2}$ assures that for a Gaussian distribution of 
observations
$\{ L_i\}$, this index is also an estimator of the standard deviation
of the distribution. Several other estimators 
of variability could be used (e.g., the
ordinary standard deviation or the light-curve amplitude), but 
the results described below are, to a large extent, independent of the
variability index employed in the analysis.

The observed variability depends not only of the quasar variability, but also
of the observational errors, and the estimation of the {\it intrinsic} 
variability $\sigma$ requires some hypothesis 
on the distribution of \{$L_i$\} and their observational errors. If we 
assume them both Gaussian distributed, the observed and intrinsic 
variabilities are related by
\begin{equation}
\sigma^2_{obs}=\sigma^2+\epsilon^2
\end{equation}
where $\epsilon^2$ 
is the mean square photometric error in luminosities.  
   
Sometimes it is more useful to work with relative variability, 
$\delta = \sigma/L$,
than with  $\sigma$ or $\sigma_{obs}$ directly. Note that $\delta$ is directly related to the
variability measured in magnitudes, $\sigma_m \simeq 0.921 ~\sigma/L$.
The relative photometric error is $\delta_\epsilon = \epsilon/L$. We
will hereafter designate the observed relative  variability by $\delta_{obs}$.
In figure 9 we display $\delta_{obs}$ as a function of absolute 
magnitude and redshift. It is also shown the median 
variability values and  the corresponding quartiles within bins with equal 
number of quasars (10). The two objects with largest variability
 in figure 9 are the high polarization quasars PKS 0537-441 
and PKS 0736+01. 

The data shown in figure 9 are consistent with a decrease
of the observed variability with both luminosity and redshift. In fact, the
Spearman rank-order correlation coefficient is 0.37 and -0.35, for the
$\delta_{obs} \times M_V$ and $\delta_{obs}
\times z$, respectively. The corresponding probabilities for the null
hypothesis of uncorrelated data sets is small, $7.7 \times 10^{-3}$
and $1.4 \times 10^{-2}$, respectively, indicating that these trends are
statistically significant.
Note that the strong correlation between $M_V$ and $z$ present in our sample
(figure 1-d) preclude us of knowing what is the real dependence of the
variability with respect to both $M_V$ and $z$.

The quasars in our sample span a large range in redshift; consequently
our observations sample  different parts of their rest-frame spectra.
This makes a direct comparison of the variability of objects 
somewhat meaningless, unless a correction for the increase of  
variability with frequency discussed in the introduction is applied. 
Next section presents a model for such a correction.

\begin{figure}
\centerline{\epsfxsize= 9cm \epsfbox{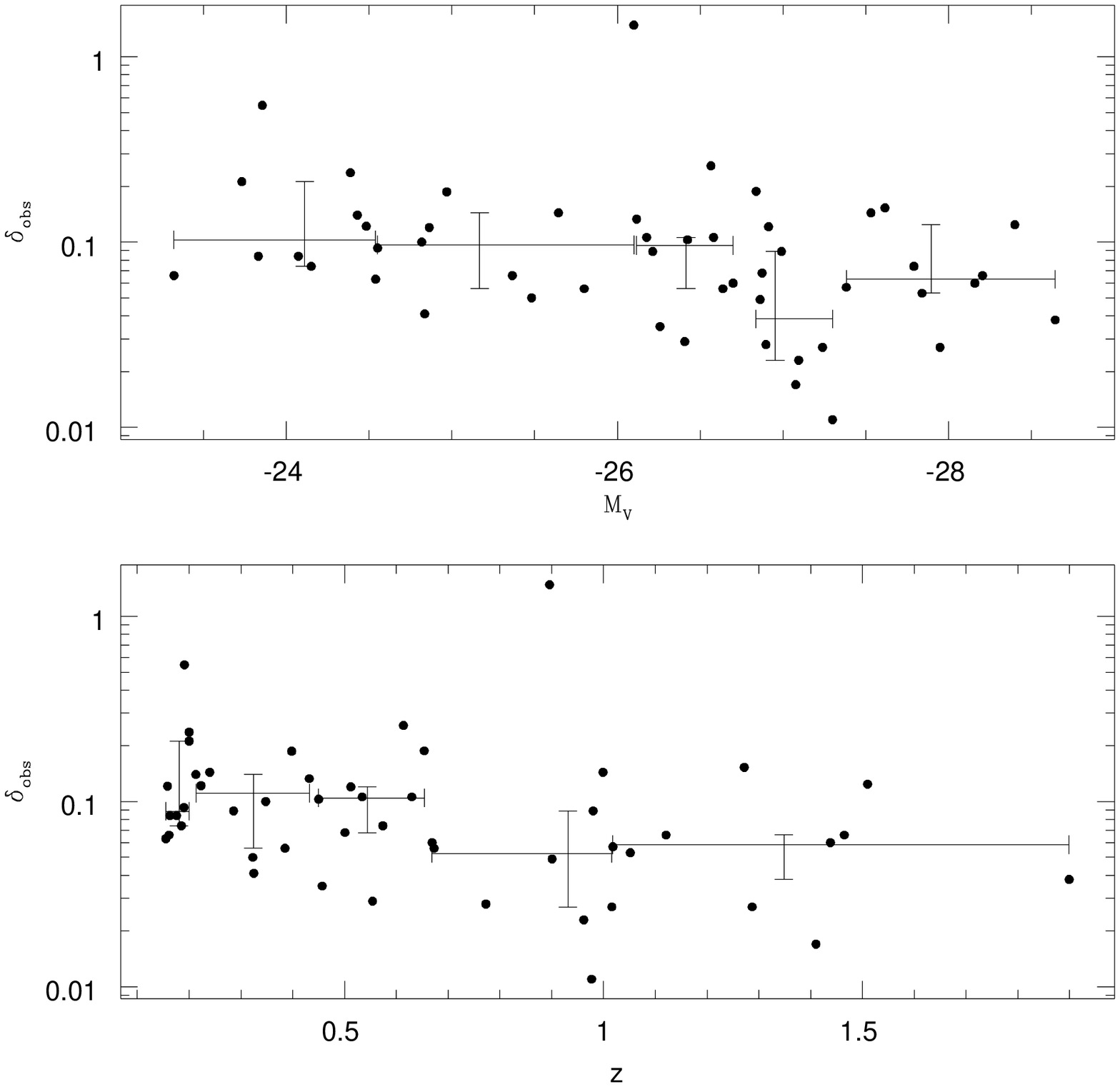}}
\caption{Observed dependences of the variability on luminosity and redshift for
our sample. Crosses represent the median values over bins of equal number of 
objects (10). The vertical bars correspond to the quartiles of the observed 
variability inside the bins, and the horizontal bars give the range in
luminosity of the objects in each bin.}
\label{fig4}
\end{figure}

\subsection{Variability $\times$ Frequency}

Let us suppose that the spectral energy distribution of a quasar 
in the optical-UV region of the spectrum is a
power-law, $f_\nu = f_* (\nu/\nu_*)^\alpha$, where $\nu$ is measured 
in the quasar rest-frame.
The increase of the variability with frequency can be interpreted as a 
hardening of the spectrum associated to an increase of luminosity. 
Following Giallongo, Trevese \& Vagnetti (1991), we assume that the
spectral index increases when the source brightens,
keeping the flux unchanged in some near-IR rest-frame frequency $\nu_*$
(Cutri et al. 1985). Edelson, Krolik \& Pike (1990) have found strong 
correlations between $\alpha_{UV}$ and the UV flux for the CfA Seyfert 1
galaxies. The same kind of correlation is exhibited by the radio-loud quasar
3C273 (Paltani \& Curvoisier 1994).
We model these spectral changes as
\begin{equation}
\alpha={\overline\alpha}+\gamma_{0} \log \left(\frac{F_{\nu_0}}
{\overline F_{\nu_0}}\right)
\end{equation}
where $\overline\alpha$ is the mean spectral index and  $\nu_0$ denotes a reference 
frequency (e.g., a filter) in the observer rest-frame.
 Despite some observational evidence that $\overline\alpha$ may be
correlated with the quasar average luminosity, we will assume that it is
constant and equal to -0.3, a value appropriate for the optical
and UV parts of the spectrum (Francis 1996, Peterson 1997). 
This value is also consistent with that we have adopted for the 
$k$-correction (section 2.1).
In Eq. (3), $\overline F_{\nu_0}$ is the average continuum flux 
integrated in a band centred on $\nu_0$, i.e.,
${\overline F_{\nu_0}} \approx f_{\nu} \Delta \nu$, where 
$\Delta \nu$ is the filter bandwidth. $\gamma_{0}$ is a parameter
that depends only of $\nu_0$. Note that the rest-frame frequency that is 
actually measured by a filter with effective frequency $\nu_0$ is
 $\nu = (1+z) \nu_0$, where $z$ is the quasar redshift.

From Eq. (3), a small change $\delta F_{\nu_0}$ 
in the quasar continuum flux produces a change in spectral index given by
$\delta \alpha = \gamma_{0} (\log e) \delta \ln F_{\nu_0}$. Since
\begin{equation}
\delta\ln f_\nu = \ln\left(\frac{\nu}{\nu_*}\right)\delta\alpha \mbox{,}
\end{equation}
we have that the monochromatic continuum flux of a quasar varies with frequency as
\begin{equation}
\frac{\delta f_\nu}{\overline f_\nu}=
\gamma_{0}\log\left(\frac{\nu}{\nu_*}\right)
\frac{\delta F_{\nu_0}}{\overline F_{\nu_0}}
\end{equation}
Let us now consider the relative variability of the flux $F_{\nu_1}$, 
measured within a band centred on a certain frequency $\nu_1$, as
\begin{equation}
\delta_{\nu_1} \equiv \frac{\delta F_{\nu_1}}{{\overline F_{\nu_1}}} \simeq 
\frac{\delta f_\nu}{{\overline f_\nu}} 
\end{equation}
where, now, $\nu = (1+z) \nu_1$ is the rest-frame frequency. Note
that in the above formula we did not take into account the line-emission correction used before to compute the absolute magnitudes of the quasars. In fact,
for objects with redshifts lower than 2.0 (our case) this correction implies differences of at most $\sim$12\% in $\delta_{V}$ (i.e., $\delta$ measured in the $V$ band), so we decided to neglect such effect. 
Then, it is easy to verify that the relative variability of a quasar
measured in two filters with effective frequencies $\nu_1$ and $\nu_2$ 
are related as
\begin{equation}
\frac{\delta_{\nu_1}}{\delta_{\nu_2}} = 
\frac{\log\left((1+z)\nu_1/\nu_*\right)}{\log\left((1+z)\nu_2/\nu_*\right)}
\end{equation} 

We have estimated $\lambda_* = c/\nu_* \approx 11400$\AA~ from data of 
Edelson, Krolik \& Pike (1990), by fitting Eq. (7) for $\lambda_1$=1450\AA~ 
and $\lambda_2$=2885\AA; we have used all objects studied by those authors
with the exception of NGC 7469, that seems to present a different 
variability pattern. Using data for 3C273 from Paltani \& Curvoisier (1994),
we obtain a similar value, $\lambda_* \approx 11250$\AA.

This very simple model for the frequency dependence of variability is indeed
able to explain the variability behaviour of the nucleus of the Seyfert
galaxy NGC 4151. For this exercise, we have used data collected in several 
spectral bands by Edelson et al. (1996). Firstly, we removed from the observed 
fluxes the light contribution of the host galaxy, accordingly to the recipe 
given in the paper. After, we fitted Eq. (7) to the relative variability
as a function of the frequency, obtaining $\lambda_* \approx$
8400\AA~ for data in the optical and UV. The fit of the model to the data
is very good, as shown in figure 10, and has a linear correlation coefficient
of -0.92. Surprisingly, including the data point corresponding to 
the intermediate X-ray emission (1.5 keV), the correlation becomes even 
stronger ($r$=-0.99) and $\lambda_*$ increases to $\sim$ 9500\AA. This
last result, however, may be somewhat fortuitous because we did not take into 
account any correction for the X-ray absorption.
Anyway, our variability model seems to fit very well the overall variability 
pattern observed in NGC 4151. 
In what follows, we adopt $\lambda_*=1$ $\mu$m as the pivot wavelength,
since the results are not
strongly dependent on the value of $\lambda_*$ (see section 4.2).
Note that near 1 $\mu$m there is also a minimum in the spectral 
energy distribution of quasars, separating the bump that extends towards
the infrared (attributed to thermal emission by dust) from the 
optical-UV Big Blue Bump.

\begin{figure}
\centerline{\epsfxsize= 9cm \epsfbox{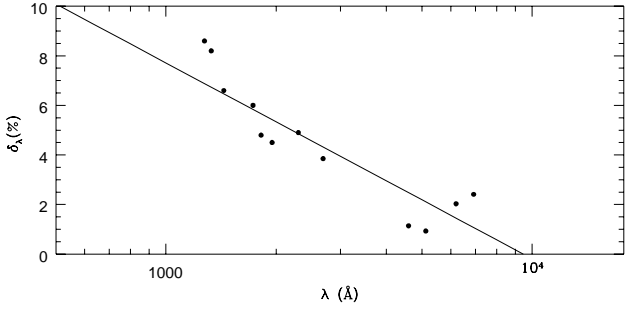}}
\vskip -4.cm
\caption{Relative flux variation (in percentage) of the NGC 4151 nucleus 
(from data of Edelson et al. 1996) as a function of the logarithm of 
wavelength. The straight line represents a simple (unweighted) linear fit.}
\label{fig4}
\end{figure}
  
Let $\delta_{obs}$ be the relative variability observed in the $V$ band  
and $\delta_{rest}$ the relative variability that we would observe in the $V$
band if the quasar was at $z=0$.
Since the $V$ band in the observer frame corresponds
to a frequency $(1+z)\nu_V$ in the rest-frame of a quasar at redshift $z$
(where $\nu_V$ is the $V$ band characteristic frequency), we have that
\begin{equation}
\frac{\delta_{rest}}{\delta_{obs}} = \frac{\log(\nu_V/\nu_*)}
{\log\left((1+z)\nu_V/\nu_*\right)}
\end{equation}
Note that the relative variability 
is independent of the $k$-correction.

These expressions can be used to transform the observed variability of the
quasars of our sample to rest-frame variability. For instance,
Eq.(8) can be rewritten numerically as
\begin{equation}
\delta_{rest}^2 = 
\left[\frac{0.260}{\log\left(1.818(1+z)\right)}\right]^2 \times
\left(\delta_{obs}^2 - \delta_\epsilon^2\right)
\end{equation}
where we have included the effect of observational errors in quadrature,
accordingly to Eq. (2). 

The effect of the correction is to reduce the observed variability.
For the most distant object in our sample ($z$=1.899) this reduction is 
36\%. 

The application of this correction decreases slightly the scattering in
the relations between relative variability and absolute magnitude or redshift.
The Spearman correlation coefficients ($r_S$) calculated with
data corrected by the measurement errors and frequency dependence are
0.38, for $\delta_{rest}^2 \times M_V$, and -0.36, for 
$\delta_{rest}^2 \times z$, to be compared with the values before the 
frequency dependence correction, 0.25 and -0.22, respectively. 

\section{Is the variability Poissonian?}

\subsection{A Simple Poissonian Model}

The non-periodicity of pulses in the light curves of quasars and 
Seyfert galaxies, as well as the observed anti-correlation between their 
variability and luminosity can be viewed as an evidence favouring the 
sub-units model. In the Starburst model (Terlevich et al. 1992), for example, 
the variability is produced by supernova explosions and the interaction of 
their ejecta with a high-density circumstellar medium. The superposition of 
random individual pulses would be responsible by the Poissonian character of 
this model. Even not being traditionally faced this way, the accretion
disc model itself could accommodate a Poissonian behaviour, as long as the 
instabilities that would cause the variability act as independent events 
(Cid Fernandes, Aretxaga \& Terlevich 1996). The same may be expected in
models where the variability is produced by external effects, if these
too are independent of each other.

Let us consider here the Poissonian model in its simplest version (e.g., 
Paltani \& Courvoisier 1997 or Cid Fernandes 1995). The light curve of a quasar is described as a 
superposition of random pulses over a  component of constant luminosity 
$L_{c}$, and depends only on three parameters related to the pulses: their
rate ($n$), energy ($E$), and effective time-scale ($t_p$).
If $l(t)$ is the typical luminosity profile of a pulse, we may define
\begin{equation}
E \equiv \int l(t) dt
\end{equation}
and
\begin{equation}
t_p \equiv \frac{\left[\int l(t) dt \right]^2}{\int l(t)^2 dt}
\end{equation}
Then, it can be shown that the average luminosity of a quasar is
\begin{equation}
L=L_c+L_{var}=L_{c}+n E
\end{equation} 
and that its variance is given by
\begin{equation}
\sigma^2=\frac{E^2n}{t_p}=\frac{E L_{var}}{t_p},
\end{equation}
where $L_{var}$ is the variable component of the luminosity.

Let $f_{var}$ be the fraction of the average luminosity due to the variable 
component: $f_{var}=L_{var}/L$. Then, the variance of the quasar luminosity
can be written as
\begin{equation}
\sigma^2=\frac{E^2n}{t_p}=\frac{E f_{var} L}{t_p},
\end{equation} 
and the relative variability is 
\begin{equation}
\delta=\frac{\sigma}{L}=\left(\frac{f_{var}E}{L~ t_p}\right)^{1/2}
\end{equation}
Therefore, if $f_{var}$, $E$ and $t_p$ are independent of the luminosity, 
$\sigma$ and $\delta$ must be proportional to $L^{1/2}$ and $L^{-1/2}$,
respectively. 
We will call this scenario as the simple Poissonian model. In this model,
quasars differ from each other only by their rate of events, $n$.

\subsection{Testing the Poissonian Hypothesis}

As mentioned before, due to the strong correlation present in the luminosity
vs. redshift diagram, we are unable to obtain  with our sample the joint 
dependence 
of the variability on luminosity and redshift. So, in order to test the 
simple Poissonian model, we will neglect any dependence of variability on 
redshift. This assumption is justified by results of some previous
variability studies, that
have not found any significant dependence of the variability vs. luminosity 
anti-correlation with redshift (e.g., Hook et al. 1994; Cristiani et al. 1996).

In order to verify whether our data is consistent with the Poissonian 
scenario, we model the corrected relative variability $\delta_{rest}$ as
\begin{equation}
\delta_{rest}^2= \frac{a}{L^{b}}
\end{equation}
where $a$ and $b$ are free parameters. In the simple Poissonian model the
expected value of $b$ is 1. 
We will work with $\delta_{rest}^2$ instead of $\delta_{rest}$
because for some objects the observed variability is smaller than
the observational errors, leading to negative values in the right side
of  Eq.(9). Our approach avoids discarding these data points. 
 
The parameters $a$ and $b$ were obtained with a ``robust" version of 
$\chi^2$. Let us define the residual $u_i$ of the variability of the
$i$-th quasar, weighted by its relative observational error, as
\begin{equation}
u_i=\frac{\left(\delta_{rest,i}^2-a/L_i^{b}\right)}{\delta_{\epsilon,i}^2}
\end{equation}
In traditional $\chi^2$, the parameters $a$ and $b$ may be determined 
by minimizing the mean 
value of the set $\{u_i^2\}$. Here we estimate these parameters
 by minimizing the median of $\{|u_i|\}$.
This approach is less sensitive to the variability behaviour of objects with 
extreme properties, like the high polarization quasars and the objects with 
very low variability.  The uncertainties in the parameters were obtained by
bootstrap re-sampling of the observed data sets (Feigelson \& Babu 1992).

The results of the application of this method to our data are shown in
Table 3. We have obtained $ a = 0.05 \pm 0.25$  
and $b = 1.20 \pm 0.58$ (for $L_V$ in units of 10$^{10} L_\odot$). 
The large values of the errors reflect the large spread of quasar variability
at a given luminosity, as can be seen in figure 9 or in
figure 11, where we plot 
$\delta_{rest}$ as a function of the luminosity of the objects, both in linear
and logarithmic scales.
The continuous line represents the best fit for all the sample. From the bottom plot in figure 11, it can be seen that excluding from the analysis objects for which $\delta_{rest}^2$ is negative would lead to a shallower slope ($b=$0.69$\pm$0.36 instead of $b=$1.20$\pm$0.58) for the variability versus luminosity relationship.

A very interesting result is that the best-fit value of $b$ is always very near the
value $b=1$ expected in the simple Poissonian model. As shown in Table 3,
even for the radio-quiet and radio-loud sub-samples we have obtained values
of $b$ consistent, within the errors, with the simple Poissonian model. 
The dashed line in figure 11 represents the best-fit solution for the
simple Poissonian model. It is worth noting that the $a$ and $b$ values obtained from this analysis are rather insensitive to changes in the pivot wavelength $\lambda_*$. For instance, modifying $\lambda_*$ by 1000\AA~ leads to changes much less than 1$\sigma$ in the best-fit values of $a$ and $b$.

If we assume that the quasar variability may indeed be described by the 
simple Poissonian model, then the only free parameter now, $a$, can be
estimated  by minimizing the median of $\{|u_i|\}$ with $b=1$. 
These results are also shown in Table 3. 
In this case the
parameter $a$ can be used to constrain the energy of the individual pulses, 
$E$, because
\begin{equation}
 E=\frac{a \times t_p}{f_{var}} 
\end{equation}
Assuming that $f_{var}$ is in the range 0.5-1.0 (Cid Fernandes, 
Aretxaga \& Terlevich 1996 adopted a lower limit of 0.3), and a 
typical pulse time-scale $t_p$ of 1.5 to 3.0 years (see, for example, 
the structure function analysis of Hook et al. 1994, and the discussion 
in Cid Fernandes, Aretxaga \& Terlevich 1996), we get 
individual energy pulses falling in to the interval 
$2.5 \times 10^{49} - 1.5 \times 10^{50}$ erg in the rest-frame $V$ band.
We can compare these results with those obtained in the analysis
of Cid Fernandes, Aretxaga \& Terlevich (1996). For quasars, we may
assume that $L_V \sim 0.66 \times L_B$ (for $\alpha=-0.3$). Then, the
energy range found by those authors is 
$1.0 \times 10^{50} \simless E \simless 4.1 \times 10^{51}$ erg 
in the $V$ band. This energy interval is not inconsistent with our results.
A recent study of the bright quasar 3C273 by Paltani, Courvoisier \& Walter 
(1998) indicates that its pulse energy in the optical-ultraviolet spectral 
range is $\sim3 \times 10^{51}$ erg. The corresponding energy in the $V$ band 
is about one order of magnitude smaller, that is also consistent with our 
results.

\newpage
\begin{table}
\caption{Parameters obtained from the robust fits, 
for all data and for the radio class sub-samples. We present results assuming
$b=1$ (simple Poissonian model), as well as for the case where $a$ and $b$
are free parameters.  The symbol $\#$ refers to the number of data points 
used in the fit. The parameters were calculated with $V$ luminosities 
in units of $10^{10} L_\odot$.}
\begin{tabular}{lccc}
& $a$ & $b$ & number \\
All Sample & 0.050$\pm$0.254 & 1.20$\pm$0.58 & 50 \\
           & 0.030$\pm$0.017 & 1             & 50 \\
           &               &               &    \\
Radio-Loud & 0.130$\pm$0.395 & 1.43$\pm$0.50 & 35 \\
           & 0.040$\pm$0.054 & 1             & 35 \\
           &               &               &    \\
Radio-Quiet & 0.020$\pm$0.150 & 1.22$\pm$0.81 & 15  \\
            & 0.014$\pm$0.010 & 1             & 15  \\
\end{tabular}
\end{table}

\begin{figure}
\centerline{\epsfxsize= 9cm \epsfbox{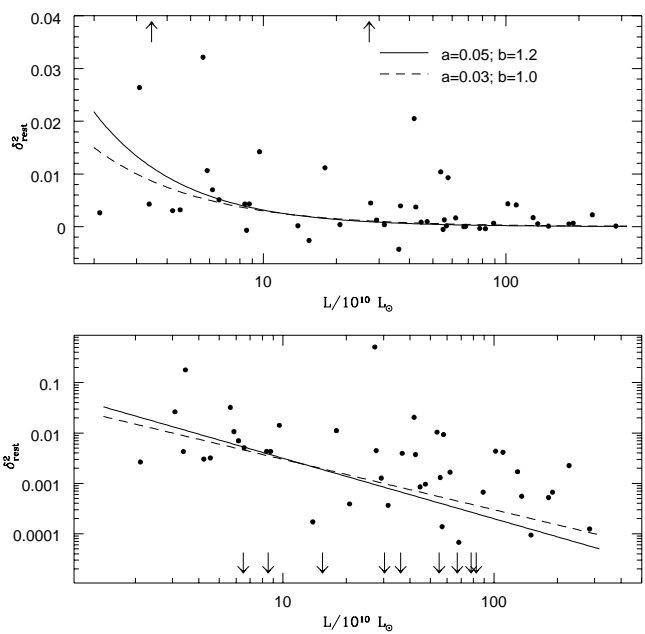}}
\caption{Rest-frame variability as a function of luminosity for our sample.
The continuous line represents the best-fit result assuming $a$ and $b$ as
free parameters, whereas the dashed line is the best-fit for the simple
Poissonian model ($b=1$). The fittings correspond to the ``all sample'' 
parameters in table 3.}
\label{fig4}
\end{figure}

\section{Variability of radio-loud and radio-quiet quasars}

Radio-loud and radio-quiet quasars (hereafter in this section 
RL and RQ, respectively)  
present almost the same continuum spectral characteristics, 
with the exception of their radio emission, where they differ by about 3 orders of 
magnitude. It has been observed a trend for RL
be more variable than RQ (Pica \& Smith 1983), what
can be due to a contribution of the collimated components associated to the 
non-thermal emission (Netzer et al. 1996). Some studies have suggested
the existence of an intermediate class, where the small scale jets
observed in some RQ would be directed along the line-of-sight 
(Miller, Rawlings \& Saunders 1993; Kellermann et al. 1994; Blundell \& Beasley 1998). 

The median values for the variability of RL and RQ, 
corrected for the variability-frequency effect are, respectively,
$\delta_{rest}$ equal to 0.063 and 0.019.
Consequently, for our sample, the rest-frame relative variability of 
RL is about 3 times larger than for RQ. About the same factor is found for
the parameter $a$ in the simple Poissonian model, as shown in Table 3.
However, RL in our sample tend to be more luminous and have higher
redshift than RQ: the median $M_V$ is -26.6 and -25.4 for RL and RQ,
respectively, and the corresponding median redshifts are 0.61 and 0.39.
According to our previous discussion, the relative variability decreases with
increasing luminosity, and, given the differences of the median values of the
luminosities of the objects in these two classes, the real difference in the
corrected relative variability of RL and RQ should be even larger.
Actually, RL tend to be more variable than RQ over 
all magnitudes and redshifts,
as can be seen in figure 12, where we plot median variability values 
calculated in bins with equal number of objects. 
Our observations then confirm earlier results obtained by Pica \& Smith (1983).

\begin{figure}
\centerline{\epsfxsize= 9cm \epsfbox{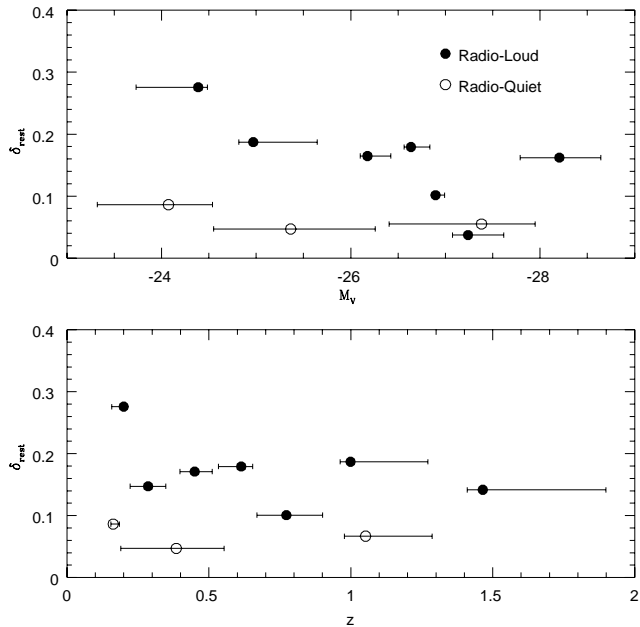}}
\caption{Median rest-frame variability as a function of 
absolute magnitude and redshift for radio-loud (filled circles)
and radio-quiet (open circles) quasars. 
The horizontal lines give the interval of
$M_V$ or $z$ corresponding to each bin.}
\label{fig4}
\end{figure}

\section{Summary and Conclusions}

In this paper we have presented the first results of a photometric
monitoring of
quasars that have been conducted at CNPq/Laborat\'orio Nacional de Astrof\1sica,
in Brazil, since 1993. The observations discussed here were done in the $V$ 
band and span the period between March 1993 and July 1996. 
This set of observations has good photometric accuracy
($\sim0.019$ $V$ mag), achieved through differential photometry with CCD 
detectors, what allows the detection of low levels of variability.

The relative variability observed in the $V$ band, $\delta_{obs}$, 
decreases with increasing luminosity and redshift. However, our
observations in the $V$ band sample different rest-frame frequencies
for quasars at different redshifts, and it is convenient to convert all 
observations to a single rest-frame frequency (e.g., that of the $V$
band) in order to compare the variability properties of these objects.
In computing $\delta_{rest}$, the relative variability expected in the
rest-frame $V$ band of the quasars, we have taken into account a model
for the increase of the variability with frequency, assuming that
an increase in luminosity corresponds to a hardening of the spectrum.
This model turned out to be consistent with the nuclear variability observed
during multi-wavelength observations of the Seyfert galaxy NGC 4151 
(Edelson et al. 1996). 
 
We have noticed that correcting the relative variability by applying the model
above decreases the scatter in the anti-correlation of $\delta$ with
$z$ or $L$, increasing slightly the significance of the relations between these
quantities. Such anti-correlations are in agreement with results of other 
studies, like Hook et al. (1994) and Cristiani et al. (1996). 
Note that the strong correlation in the $M_V \times z$ diagram for the
objects in our sample preclude us of disentangling the real dependences 
of the variability on these two parameters. 

We have verified that the relative variability $\delta_{rest}$ 
tends to decrease with the luminosity in agreement with the expectations of 
a simple Poissonian model, where the light curves are assumed to be
the result of a random superposition of pulses of same energy. 
Characteristic energies for individual pulses 
derived with such a model are consistent with those predicted by the nuclear
starburst scenario (e.g., Cid Fernandes, Aretxaga \& Terlevich 1996). 
Assuming that there are no differences in $n$, $f_{var}$ and $t_p$ for both classes, the pulse energies for the radio-loud
objects are larger than for radio-quiet objects by almost one order of
magnitude.
The radio-loud objects in our sample tend to be more variable than 
radio-quiet objects. 

\section*{ACKNOWLEDGEMENTS}

This work benefited from the financial support pro\-vi\-ded by the Brazilian
agencies FAPESP and CNPq. We are grateful to Eduardo Cypriano and H\'ector Cuevas for helping in the observations, and to Roberto Cid Fernandes
and Sueli Viegas for fruitful discussions.


\begin{thebibliography}{1}
\bibitem[AL 1976]{al76}
Allen C.W., {\it Astrophysical Quantities}, 3th Edition, 1976
\bibitem[ACT 1997]{act97}
Aretxaga I., Cid Fernandes R., Terlevich R.J., 1997, MNRAS, 286, 271
\bibitem[BE 1990]{be90} 
Bessel M.S., 1990, PASP, 102, 1181
\bibitem[BS 1994]{bs94}
Borgeest U., Schramm K.J., 1994, A\&A,284,764
\bibitem[BB 1998]{bb98}
Blundell K.M., Beasley A.J., 1998, MNRAS, to be published
\bibitem[CF 1995]{cf95}
Cid Fernandes R., 1995, PhD thesis, Cambridge, England
\bibitem[CAT 1996]{cat96}
Cid Fernandes R., Aretxaga I., Terlevich R., 1996, MNRAS, 282, 1191
\bibitem[CZM 1993]{czm93}
Cimatti A., Zamorani G., Marano B., 1993, MNRAS, 263, 236 
\bibitem[CTLAA 1996]{ctlaa96} 
Cristiani S., Trentini S., La Franca F., Andreani P., Aretxaga I., 1996,
A\&A, 306, 395
\bibitem[CVA 1997]{cva97} 
Cristiani S., Trentini S., La Franca F., Andreani P., 1997,
A\&A, 321, 123
\bibitem[CVA 1990]{cva90} 
Cristiani S., Vio R., Andreani P., 1990, AJ, 100, 56
\bibitem[CWRL 1985]{cwrl85}
Cutri R.C., Wisniewski W.Z., Rieke G.H., Lebofsky M.J., 1985, ApJ, 296, 423
\bibitem[EKP 1990]{ekp90} 
Edelson R., Krolik J., Pike G., 1990, ApJ, 359, 86
\bibitem[ED 1996]{ed96}
Edelson et al., 1996, ApJ, 470, 364
\bibitem[FB 1992]{fb92}
Feigelson E.D., Babu G.J., 1992, ApJ, 397, 55
\bibitem[]{fr}
Francis P., 1996, PASA, 13, 212
\bibitem[GTV 1991]{gtv91}
Giallongo E., Trevese D., Vagnetti F., 1991, ApJ, 377, 345
\bibitem[GMKNS 1999]{gmkns99}
Giveon U., Maoz D., Kaspi S., Netzer H., Smith P.S., 1999, astro-ph/9902254
\bibitem[GSW 1995]{gsw95} 
Gopal-Krishna, Sagar R., Wiita P.J., 1995, MNRAS, 274, 701
\bibitem[GR 1982]{gr92}
Graham J.A.,1982, PASP, 94, 244
\bibitem[HMBI 1994]{hmbi94}
Hook I.M., MacMahon M.G., Boyle B.J., Irwin M.J., 1994, MNRAS, 268, 305
\bibitem[KSSGS 1994]{kssgs94}
Kellermann, K.I., Sramek R.A., Schmidt M., Green R.F., Shaffer D.B., 1994,
AJ, 108, 4
\bibitem[KBBY 1991]{kbby91}
Kinney A.L., Bohlin R.C., Blades J.C., York D.G., 1991, ApJS, 75, 645
\bibitem[LL 1984]{ll84}
Lloyd C., 1984, MNRAS, 209, 697
\bibitem[MRS 93]{mrs93}
Miller P., Rawlings S., Saunders R., 1993, MNRAS, 263, 425
\bibitem[NHLAB 1996]{nhlab96}
Netzer H., Heller A., Loinger F., Alexander T., Baldwin J.A., Wills B.J.,
Han M., Frueh M. Higdon J.L., 1996, MNRAS, 279, 429
\bibitem[PC 1994]{pc94}
Paltani S., Courvoisier T.J.-L., 1994, A\&A, 291, 74
\bibitem[PC 1997]{pc97}
Paltani S., Courvoisier T.J.-L., 1997, A\&A, 323, 717
\bibitem[PCW 1998]{pcw98}
Paltani S., Courvoisier T.J.-L., Walter R., 1998, A\&A, 340, 47
\bibitem[PS 1983]{ps83}
Pica A.J., Smith A.G., 1983, ApJ, 272, 11
\bibitem[]{pet97}
Peterson B.M., 1997, {\it An Introduction to Active Galactic Nuclei},
Cambridge University Press
\bibitem[RE 1984]{re84}
Rees M.J., 1984, ARA \& A, 22, 471
\bibitem[TTFM 1992]{ttfm92}
Terlevich R., Tenorio-Tagle G., Franco J., Melnick J., 1992, MNRAS, 255, 713
\bibitem[TKM 1994]{tkm94}
Trevese D., Kron R. G., Majewski S.R., Bershady M.A. \& Koo D.C., 1994, ApJ, 433, 494
\bibitem[UWW 1976]{uww76}
Uomoto A.K., Wills B.J., Wills D., 1976, AJ, 81, 905
\bibitem[VCV 1987]{vcv87}
V\'eron-Cetty M.P., V\'eron P., 1987, ESO Scientific Report, 3rd edition
\bibitem[vcv 1993]{vcv93}
V\'eron-Cetty M.P., V\'eron P., 1993, ESO Scientific Report, 6th edition

\end{thebibliography}
\end{document}